\documentclass{PoS}

\title{Theory overview}

\ShortTitle{Theory Overview}

\author{\speaker{Alexander Lenz}\\
        IPPP, Durham University, United Kingdom\\
        E-mail: \email{alexander.lenz@durham.ac.uk}}

\abstract{We set the scene for theoretical issues in charm physics that were discussed
          at CHARM 2016 in Bologna. In particular we emphasize the importance of improving our understanding
          of standard model contributions to numerous charm observables and we discuss also possible tests
          of our theory tools, like the Heavy Quark Expansion via the lifetime ratios of $D$-mesons.} 

\FullConference{VIII International Workshop On Charm Physics\\
		5-9 September, 2016\\
		Bologna, Italy}

\begin{document}

\section{Introduction}
First of all I would like to thank the organisers for giving me the possibility of opening this conference and for 
choosing such an inspiring venue, as the Convent of San Domenico. Besides having a long intellectual tradition, 
going back to scholars like Albertus Magnus, Meister Eckart and Thomas Aquinas,
these sages had an interesting approach of dealing with different-minded people, which we might re-consider nowadays, 
in particular when dealing with 
politicians, who endanger the scientific progress with nationalistic and racist attitudes.
\\
The idea of this talk was not to anticipate any of the results that will be shown later in the conference; 
they will be covered by the individual talks and by the theory summary \cite{CHARM01}. Moreover, to give
a further indication of the broad range of theoretical activities in this field, I mostly quote results from
groups that are not present at CHARM 2016. Marco will present the corresponding experimental introduction
\cite{CHARM02}, including historical remarks.
The topics discussed in Bologna were grouped into the following sessions:
{\it Heavy Ions},
{\it Multi-body hadronic decays and amplitude analysis},
{\it Leptonic, semi-leptonic and rare decays (CKM elements)},
{\it Charm Baryon decays},
{\it Charmonium and Exotics, production and spectroscopy},
{\it CP violation, Mixing and non-leptonic decays},
{\it Open Charm production and spectroscopy} and
{\it Future Prospects}.
In order to give a systematic, brief overview of our field and its aims
as well as to point out some directions, that might be
important for the future development of our field
I have chosen to re-categorise these topics into:
\begin{enumerate}
\item What is special about Charm?
      \begin{itemize}
      \item the mass of the  charm quark is neither heavy nor light; thus it is questionable if theory tools 
            like the Heavy Quark Expansion (\cite{HQE} or \cite{HQEreview} for a review) or factorisation 
            \cite{QCDfac} work in the charm system.
     \item  the charm system is subjected to severe GIM \cite{GIM} cancellations. This standard model 
            peculiarity might not be present in extensions of it. Moreover such a strong numerical effect might also
            overshadow tests of the applicability of theory tools.
     \item  due to different couplings and parameters the investigation of charmed hadrons is complementary to the 
            well-studied Kaons and B-mesons. 
     \item  finally a very pragmatic motivation for studying the charm system: we have a huge amount of charm data, 
            e.g. from LHCb or BESIII (see e.g. \cite{Gersabeck:2015wna}) and there is much more to come from
            Belle II \cite{CHARM03}, the LHC-upgrade \cite{CHARM04}, BESIII \cite{CHARM05} and PANDA \cite{CHARM06}.
     \end{itemize}
\item Understanding of QCD: having a quantitative understanding of hadronic effects in the charm sector is
      absolutely crucial for any conclusions about possible new physics effects in the charm system; a very instructive 
      example for that statement is the $\Delta A_{CP}$-saga, see e.g. \cite{Lenz:2013pwa}.
    \begin{itemize}
    \item Spectroscopy, exotics: the theoretical description and understanding of bound-states including charm quarks
          is a very active research field, in particular since exotic states like penta-quarks have been
          established experimentally \cite{Aaij:2016phn,Aaij:2016ymb}; we had nine dedicated theory talks in Bologna:
          \cite{CHARM07}, 
          \cite{CHARM08}, 
          \cite{CHARM09}, 
          \cite{CHARM10}, 
          \cite{CHARM11}, 
          \cite{CHARM12}, 
          \cite{CHARM13}, 
          \cite{CHARM14}, 
          \cite{CHARM15}.
    \item Charm contributions in heavy ion physics might shed more light into the nature of the 
          quark-gluon plasma and thus on cosmology, see e.g.
          \cite{CHARM16}, 
          \cite{CHARM17}, 
          \cite{CHARM18}, 
          \cite{CHARM19}, 
          \cite{CHARM20} 
          at this conference.
    \item Charm production is described by perturbative QCD, we had two presentations at CHARM 2016: 
           \cite{CHARM21}, 
           \cite{CHARM22}. 
    \item Leptonic and semi-leptonic decays have the simplest possible hadronic structure, they depend
          on non-perturbative decay constants and  form factors, which can be determined on the lattice or 
          with sum rules. The lattice progress was described in
          \cite{CHARM23}. 
          These decays seem to be good candidates for new physics searches
          \cite{CHARM24}, 
          \cite{CHARM25}. 
    \item Hadronic decays are considerably more difficult to be described in theory, thus it is not clear whether
          tools like QCD factorisation give us any insight and one has to use assumptions
          like $ SU(3)_F$-symmetry to make predictions. This topic was also intensively discussed at CHARM 2016:
          \cite{CHARM26}, 
          \cite{CHARM27}, 
          \cite{CHARM28}, 
          \cite{CHARM29}, 
          \cite{CHARM30}. 
    \item Mixing: a naive application of the theory tools that work well in the $B$-system to the $D$-system 
          gives results that are orders of magnitudes away from the experimental result.
          Here progress is urgently needed to make use of the relatively precise data - mixing was discussed in Bologna in
          \cite{CHARM31}, 
          \cite{CHARM32}. 
          Lattice might turn out to yield promising results for $D$-mixing on a longer time scale
          \cite{CHARM33}; 
          on a shorter time scale a precise theoretical investigation of charm lifetimes \cite{Lenz:2013aua}, 
          which is doable with
          current lattice technology, could shed light into the convergence properties of the HQE in the charm-system.
    \end{itemize}
\item Determination of Standard Model parameters:
     \begin{itemize}
     \item The CKM elements $V_{cs}$ and $V_{cd}$ are among the least well known mixing parameters; 
           their measurement provides a test of the unitarity of the CKM matrix. The impact of charm physics to the 
           CKM fit was discussed in \cite{CHARM34}.
     \item The precise value of the charm quark mass $m_c$ is needed for e.g. precision predictions in the 
           $b$-quark sector. This topic was not discussed in Bologna.
     \end{itemize}
\item Search for new physics effects in the charm sector are complementary to many of the current indirect search strategies:
     \begin{itemize}
     \item D-meson decays (leptonic, semi-leptonic and hadronic ones) were discussed in that respect by
           \cite{CHARM24,CHARM25}. 
           If new physics particles are heavy then our theory tools could work again well for the new
           contributions; unfortunately it is still very problematic to estimate the size of the Standard Model part.
     \item A study of the Higgs-Yukawa coupling ($H \to  c \bar{c}$) was suggested several times in the recent literature, 
           see e.g.
           \cite{Perez:2015aoa,Bishara:2016jga,Botella:2016krk,Brivio:2015fxa,Koenig:2015pha}.
     \item There are almost no studies of dark matter candidates that couple to the up-type quark sector, see \cite{charmDM}.
     \item Indirect charm contributions to quantities that are very sensitive to new physics effects
           are currently studied on the lattice, e.g. $g-2$ \cite{Wittig}, $\epsilon_K$ \cite{Sachrajda,Feng}.
     \end{itemize}
\item Our understanding of Quantum Mechanics might be improved by quantum coherent charm measurements; this was
      discussed in \cite{CHARM35}.
\end{enumerate}

Due to a limitation in space we will not discuss all these topics in the proceedings.

\section{What is special about Charm?}
    The masses of the charm and the bottom quarks are now very well determined. In \cite{Maezawa:2016vgv} values of
    \begin{eqnarray}
    \bar{m_c}( \bar{m_c}) & = & 1.267(11) \, \rm GeV \, ,
    \\
    \bar{m_b}( \bar{m_b}) & = & 4.183(83) \, \rm GeV 
    \end{eqnarray}
    were obtained, using lattice QCD. 
    The large value of the bottom quark mass enables an expansion of inclusive decay rates in the inverse of this value
    \cite{HQE}:
    \begin{equation}
    \Gamma = \Gamma_0 + \frac{\Lambda^2}{m_b^2} \Gamma_2 
                      + \frac{\Lambda^3}{m_b^3} \Gamma_3
                      + \frac{\Lambda^4}{m_b^4} \Gamma_4 + \dots  \, ,
    \end{equation}
    where $\Lambda$ is a hadronic scale. The convergence of the HQE in the bottom sector was proven \cite{Lenz:2012mb}
    by the agreement of experiment
    \cite{Amhis:2014hma}  and theory \cite{Artuso:2015swg} 
    (based on \cite{Lenz:2006hd,Beneke:2003az,Ciuchini:2003ww,Beneke:2002rj,Dighe:2001gc,Beneke:1998sy,Beneke:1996gn})
    for the decay rate difference $\Delta \Gamma_s$ in the neutral $B_s^0$-system.
    \begin{equation}
    \Delta \Gamma_s^{\rm HFAG} = (0.083 \pm 0.006) {\rm ps}^{-1} \, ,
    \hspace{1cm}
    \Delta \Gamma_s^{\rm SM} = (0.088 \pm 0.020) {\rm ps}^{-1} \, .
    \end{equation} 
    The charm quark mass is roughly a factor of three smaller than the bottom-quark mass and thus much closer
    to the hadronic scale  $\Lambda_{QCD}$. Hence it is questionable if the HQE is still converging, even if it does
    not seem unreasonable a priori.
    The experimental values for the mixing observables in the charm sector read \cite{Amhis:2014hma}:
    \begin{equation}
       x_D^{\rm Exp.} = \frac{\Delta M_D     }{  \Gamma_D} = (0.37 \pm 0.16) \cdot 10^{-2} \, ,
       \hspace{1cm}
       y_D^{\rm Exp.} = \frac{\Delta \Gamma_D}{2 \Gamma_D} = (0.66^{+0.07}_{-0.10}) \cdot 10^{-2} \, .
   \end{equation}
   To test the applicability of the HQE we simply adopt the formulae from the $B$-sector to the $D$-mesons
   \cite{Bobrowski:2010xg} (including $\alpha_s(m_c)$- and $\Lambda / m_c$-corrections)
   \begin{eqnarray}
      y_D^{\rm HQE} & \leq & |\Gamma_{12}^D| \tau_D \, ,
      \\
      \Gamma_{12}^D & = & - \left(    \lambda_s^2         \Gamma_{12}^{ss}
                                 + 2 \lambda_s \lambda_d \Gamma_{12}^{ds}
                                 +   \lambda_d^2         \Gamma_{12}^{dd}
       \right) \, .
    \end{eqnarray}
    $\lambda_q $ denotes CKM structures and the $\Gamma_{12}^{pq}$ are the loop contributions with an internal
    $p$- and $q$-quark.
    \\
    \includegraphics[width = 0.9 \textwidth]{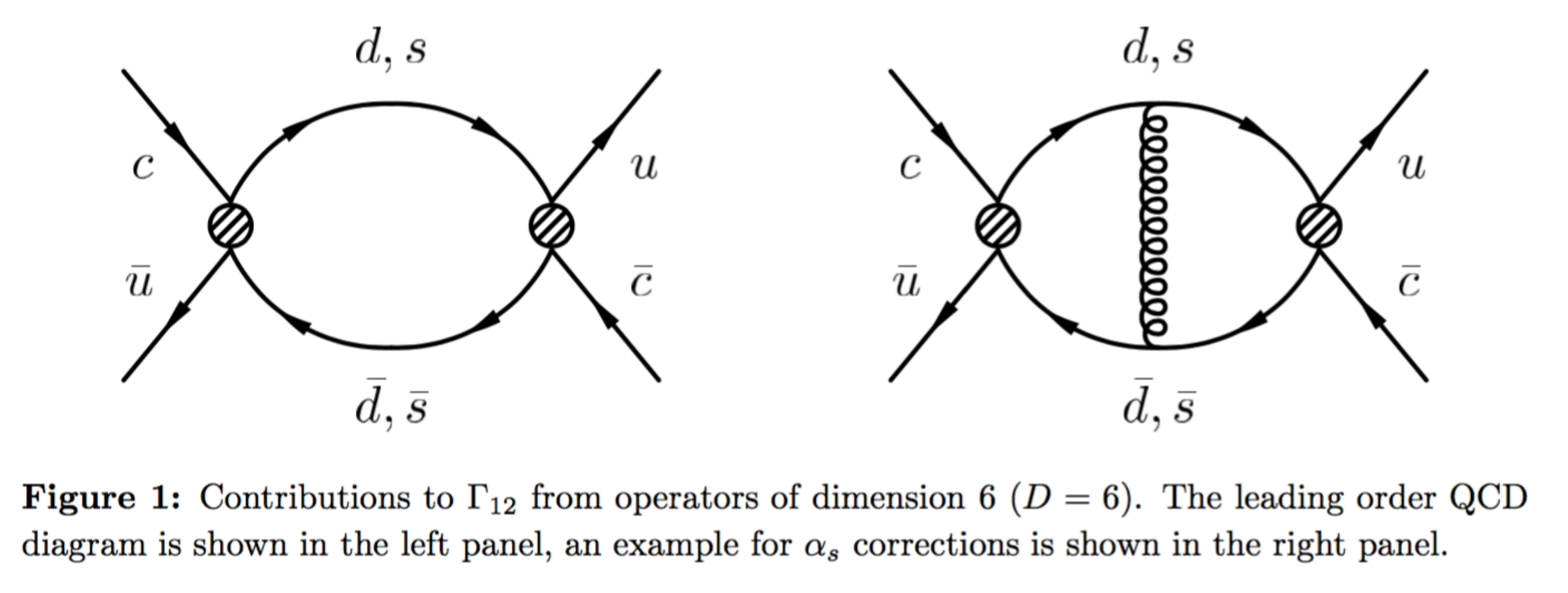}
    \\
    Considering only the $s$-quark contribution, we get
    \begin{eqnarray}
      y_D^{\rm HQE} & \supset &    -     \lambda_s^2         \Gamma_{12}^{ss} \tau_D \approx 5.6  y_D^{\rm Exp.}  \, .
    \end{eqnarray}
    Thus a single diagram gives a contribution that is larger than the experimental value. Considering now 
    all three contributions and using in addition the unitarity of the CKM matrix, we find a severe GIM cancellation
    \begin{eqnarray}
      y_D^{\rm HQE} & \approx &    -     \lambda_s^2  \left(   
     \Gamma_{12}^{ss} - 2 \Gamma_{12}^{sd} + \Gamma_{12}^{dd} \right) \tau_D \approx 1.7 \cdot 10^{-4}  y_D^{\rm Exp.}
       \, ,
    \end{eqnarray}
    pushing the HQE prediction far below the experimental value. Similar GIM cancellations appear also in rare 
    penguin induced $D$-decays. Below we will discuss implications of this severe cancellations.

\section{Understanding of QCD}
    For spectroscopy, exotics, heavy ions and charm production we refer the reader to the individual contributions
    \cite{CHARM07,CHARM08,CHARM09,CHARM10,CHARM11,CHARM12,CHARM13,CHARM14,CHARM15,CHARM16,CHARM17,CHARM18,CHARM19,CHARM20,CHARM21,CHARM22} and we concentrate here on meson decays and mixing.
    Leptonic decays, like $D_s^+ \to \mu^+ + \nu_\mu$,  posses the simplest hadronic structure, 
    which is parameterised by a decay constant $f_{D_s}$:
   \begin{equation}
   \langle 0 | \bar{c} \gamma_\mu \gamma_5 q | D_q(p) \rangle = i f_{D_q} p^\mu_{D_q} \, .
   \end{equation}
   The theoretical determination with sum rules and lattice QCD of decay constants is quite advanced, see e.g. \cite{Rosner:2015wva},
   and it agrees well with experimental measurements:
   \\
   \includegraphics[width = 0.9 \textwidth]{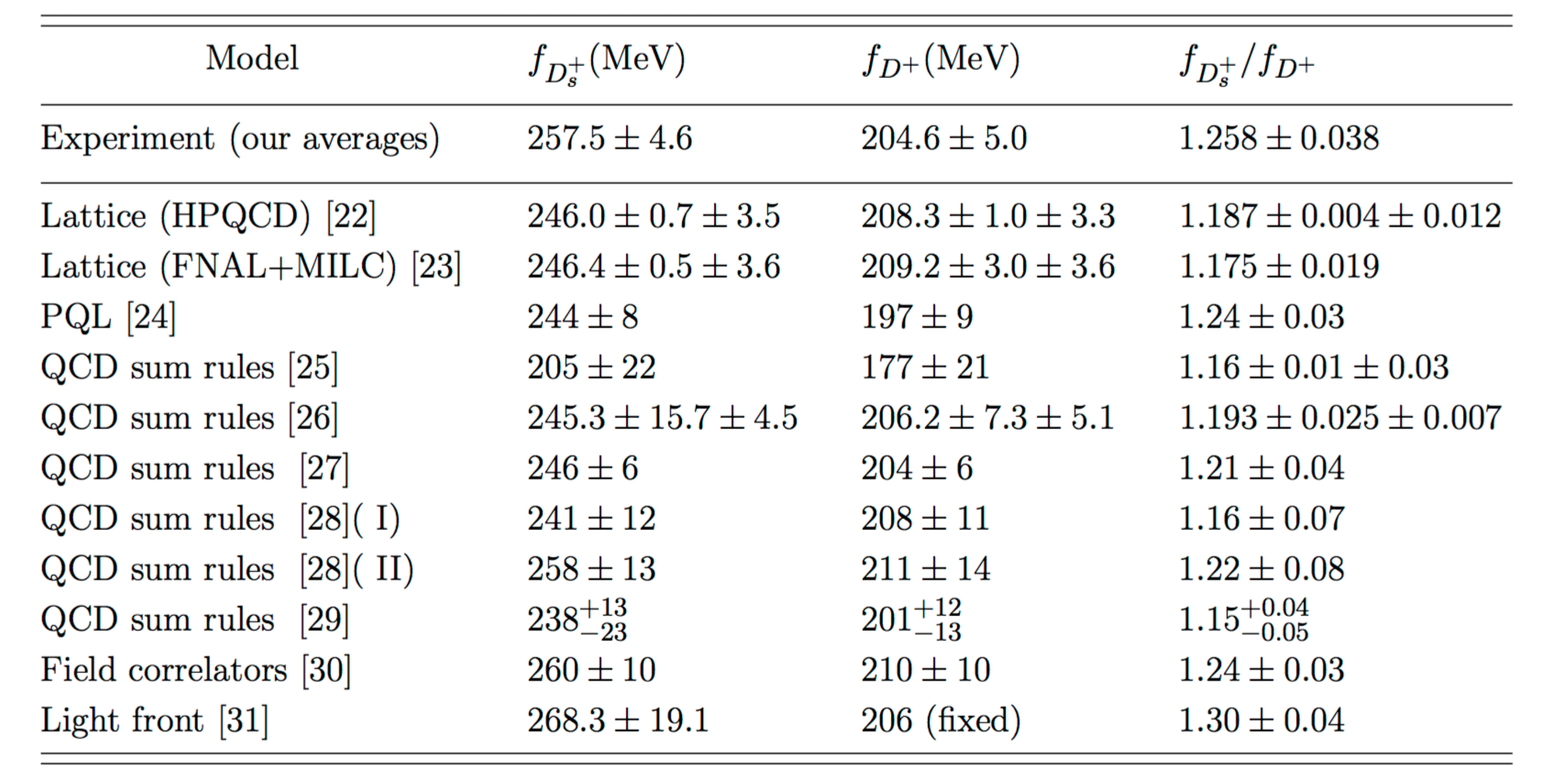}
    \begin{flushright}
    Table from Ref. \cite{Rosner:2015wva}.
    \end{flushright}
   From a hadronic point of view the next complicated class of decays are semi-leptonic decays. Here the non-perturbative 
   part is parameterised by form factors
   \begin{equation}
   \langle K | V^\mu | D \rangle = f_+(q^2) \left(p^\mu_D + p^\mu_K - \frac{M_D^2 - M_K^2}{q^2} q^\mu \right) 
                              + f_0(q^2)                           \frac{M_D^2 - M_K^2}{q^2} q^\mu \, ,
   \end{equation}
   where $f_+$ and $f_0$ denote the form factors, that depend on the momentum transfer. To get an idea of the currently achieved
   theoretical precision for decay constants and form factors we show a plot from a talk of Steven Gottlieb at LATTICE 2016
   \cite{Gottlieb}.
   \\
   \includegraphics[width = 0.9 \textwidth]{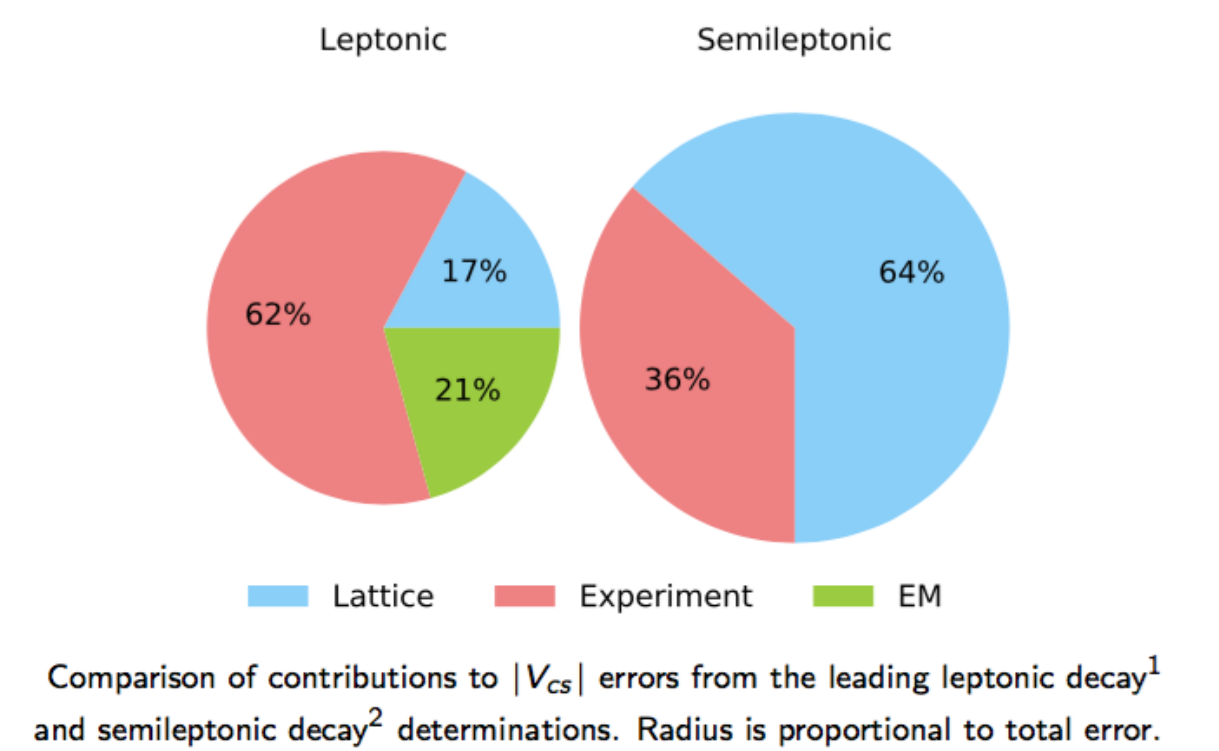}
    \\
   Even more complicated are purely hadronic decays like $D \to \pi \pi$ or $D \to K K$. It is not clear at all, that matrix 
   elements like
   \begin{equation}
   \langle \pi \pi | Q |D \rangle \, ,
   \end{equation}
   where $Q$ is a four-quark operator, factorise. To make nevertheless theoretical statements, 
   assumptions like $SU(3)_F$-symmetry are regularly used in the literature, 
   see e.g. \cite{Muller:2015lua,Gronau:2015rda,Gao:2014ena,Jung:2014jfa,Santorelli:2014kva,Hiller:2013awa,Buccella:2013tya}
   for some recent (2013 onward) references. In the long term future this problem might be solvable with lattice QCD, by an extension
   \cite{Hansen:2012tf} of methods that have shown to be very successful for hadronic Kaon decays \cite{Sachrajda}.
   \\ 
   Finally we are coming back to mixing. The off-diagonal elements of the matrix describing  the mixing of neutral $D$-mesons 
   can be expressed as
   \begin{equation}
   2 M_D \left(M_{12}^D - \frac{i}{2} \Gamma_{12}^D \right) = 
  \langle D^0| {\cal H}_{\rm eff}^{| \Delta C |= 2}| \bar{D}^0 \rangle
  + \sum \limits_n
  \frac{
  \langle D^0| {\cal H}_{\rm eff}^{| \Delta C |= 1}| n         \rangle
  \langle n  | {\cal H}_{\rm eff}^{| \Delta C |= 1}| \bar{D}^0 \rangle
  }{M_D - E_n + i \epsilon} \, .
   \end{equation}
  The first term on the r.h.s. is short distance dominated. The necessary matrix elements of the 4-quark operators
  have been determined on the lattice \cite{Carrasco:2015pra,Kronfeld}; this part is thus relatively well understood. It is also
  interesting to note that heavy new physics particles contribute predominantly to the short distance part.
  The second term on the r.h.s. is dominated by light internal particles. In the $B$-system the corresponding contribution 
  to $M_{12}$ turned out to be negligible (due to the CKM structure and the large value of the top-quark mass)
and $M_{12}$ is given to a very good approximation by the $| \Delta B = 2|$ contribution alone.
  This is not the case anymore in the $D$-system. On the other hand, $\Gamma_{12}$ is governed by on-shell intermediate particles, 
  hence only
  the   $| \Delta B,C = 1|$-parts are contributing,
  In the case of $B$-mixing it turned out to be possible to use quark hadron duality and perform successfully an expansion 
  in $\Lambda / m_b$, the Heavy Quark Expansion. In the $D$-system
  one expects the expansion parameter to be a factor three larger, which might still be ok. A related way of looking at that, 
  is to compare the remaining
  phase space in the decay channels contributing to the decay rate difference with the hadronic scale.
  \begin{eqnarray}
  M_{B_s^0} - 2 M_{D_s^{(*)}} & = & 1.43 (1.15) \, {\rm GeV} \, ,
  \\
  M_{B_s^0} - M_{J/ \psi} - M_{\phi} & = & 1.25 \, {\rm GeV} \, ,
  \\
  M_D - 2 M_K               & = & 0.88        \, {\rm GeV} \, ,
  \\
  M_D - 2 M_\pi             & = & 1.59        \, {\rm GeV} \, .
  \end{eqnarray}
  These numbers for $\Delta \Gamma_s$ and $\Delta \Gamma_D$ are quite similar, so a priori it seems not obvious that the 
  HQE does not converge in the charm system, on the other we have seen above
  that the leading terms of the HQE give results that are about four orders of magnitude below the experimental value.
  Instead of the inclusive approach which assumes quark hadron duality one can try to use the exclusive approach, where the sum over
  all possible final states into which both the $D^0$-meson and the $\bar{D}^0$-meson can decay, has to be performed. 
  This is obviously even more
  challenging than calculating only one hadronic D-meson decay. Nevertheless this approach together with several simplifying 
  assumptions was used in \cite{Falk:2001hx,Falk:2004wg} to predict roughly the experimental values of the charm mixing 
  observables. Here the next step would be to reduce the number of assumptions and make the theory predictions more realistic; 
  there are indications that on a long time scale this issue could be solved on the lattice by further developments of the methods 
  from Hansen and Sharpe \cite{Hansen:2012tf}.
  \\   
  But even, if it turns out in future  that the exclusive approach will reproduce the experimental $D$-mixing values, it is not
  clear, why the inclusive approach is failing by about four orders of magnitude, despite working so well in the $B$ sector and 
  despite having an expansion parameter that is only a factor of about three larger.
  Above we already observed that the  individual diagrams for $\Gamma_{12}^D$ give values larger than $y^{\rm Exp}$ and only the 
  combination of all three contributing diagrams is more or less vanishing due to GIM cancellations. This old observation  
  was revived recently in \cite{Jubb:2016mvq}, where it was found that a modification of individual diagrams of the order of
  $20 \%$ due to some unknown duality violating effect, would be sufficient to lift the GIM suppression so much, that the 
  experimental value of $y$ could be reproduced by the leading term in the HQE. 
  \\
  In the same spirit, it was already argued several years ago
  \cite{Georgi:1992as,Ohl:1992sr,Bigi:2000wn} that the GIM suppression might be lifted in higher orders of the HQE. If the lifting
  of the GIM suppression is more pronounced than the suppression due to $\Lambda / m_c$, then the dominant contribution of the HQE
  prediction for $\Gamma_{12}^D$ might actually stem from dimension 9 or dimension 12 contributions, see also \cite{Bobrowski:2010xg}.
  A first step in that direction was already done in \cite{Bobrowski:2012jf}, which seems to indicate that dimension nine is in fact 
  larger than the leading dimension six contribution, but unfortunately still considerably below experiment.
  \\
  Next one could continue to determine the dimension twelve contributions and thus test the above idea of the dominance of
  higher orders in the HQE. A severe limitation of this approach will, however,  be the treatment of the unknown matrix elements
  of the 8-fermion operators. Except naive factorisation there is currently  no adequate theory tool in sight.
  An alternative test of the convergence of the HQE in the charm-system would be the investigation of observables that are free of 
  GIM-cancellations. Here lifetime ratios are the prime candidates.
  The experimental value 
  of the lifetime ratio shows quite a large deviation from one
  \begin{equation}
  {\frac{\tau (D^+)}{\tau (D^0)}}^{\rm Exp}=  2.536 \pm 0.019 \, .
  \end{equation} 
  This number does, however, not necessarily point towards a $150 \%$ correction in the HQE, a $40 \%$ correction 
  could also easily do the job,
  via $(1+0.4)/(1-0.4) = 2.3$. The perturbative part of the HQE prediction for this lifetime ratio 
  is known up to next-to-leading order in QCD and in the parameter $\Lambda / m_c$
  \cite{Lenz:2013aua}; unfortunately the non-perturbative matrix elements of the arising 4-quark operators 
  \begin{eqnarray}
  Q^q   & = & \bar{c} \gamma_\mu ( 1- \gamma_5) q \cdot \bar{c} \gamma^\mu ( 1- \gamma_5) q \, ,
  \\
  Q_S^q & = & \bar{c}            ( 1- \gamma_5) q \cdot \bar{c}            ( 1+ \gamma_5) q \, ,
  \\
  T^q   & = & \bar{c} \gamma_\mu ( 1- \gamma_5) T^a q \cdot \bar{c} \gamma^\mu ( 1- \gamma_5) T^a q \, ,
  \\
  T_S^q & = & \bar{c}            ( 1- \gamma_5) T^a q \cdot \bar{c}            ( 1+ \gamma_5) T^a q 
  \end{eqnarray}
  have not yet been 
  determined. This task seems to be perfectly doable with current lattice technology. 
  Making some simplifying assumptions for the unknown matrix elements  Ref.\cite{Lenz:2013aua} obtained
  \begin{equation}
  {\frac{\tau (D^+)}{\tau (D^0)}}^{\rm HQE}=  {2.2 \pm 1.7^{\rm hadronic} }^{+0.3 \rm scale}_{-0.7} \pm 0.1^{\rm parametric} \, .
  \end{equation} 
  This result is promising, but unfortunately not conclusive, due to the huge uncertainties related to the unknown matrix elements.
  Here a lattice study could shed very valuable insight into the applicability of the HQE.
  Finally it is entertaining to note that there is still the possibility of having found new physics in $D$-mixing 
  without having noticed it yet.

\section{Determination of Standard model parameters:}
  The direct determination of the CKM elements $V_{cs}$ and $V_{cd}$ still suffers from considerable uncertainties. PDG \cite{PDG}
  quotes values of
  \begin{eqnarray}
  V_{cd}  & = & 0.225  \pm 0.008 \, , 
  \\
  V_{cs}  & = & 0.986  \pm 0.016 \, , 
  \\
  V_{cb}  & = & 0.0411 \pm 0.0013 \, . 
  \end{eqnarray}
  This leads to the following test of the unitarity of the CKM matrix
  \begin{eqnarray}
  | V_{cd}|^2 +   | V_{cs}|^2 + | V_{cb}|^2  & = & 1.024  \pm 0.032 
                                              = 1 + (0.15)^2 \pm (0.18)^2 \, . 
  \end{eqnarray}
  Thus there is still space for sizable new physics effects and the corresponding CKM values have to be determined more precise in future.
  The status quo of CKM fits will be reviewed by  Derkach \cite{CHARM34}.
  \\
  The charm quark mass $m_c$ has been determined by many groups and has obtained an impressive precision, which is necessary
  for making precise predictions in other fields, like $b$-physics, where the charm quark mass is an important input parameter. 
  An overview of lattice  determinations was shown at LATTICE 2016 by Nakayama \cite{mc}:
    \\
    \includegraphics[width = 0.9 \textwidth]{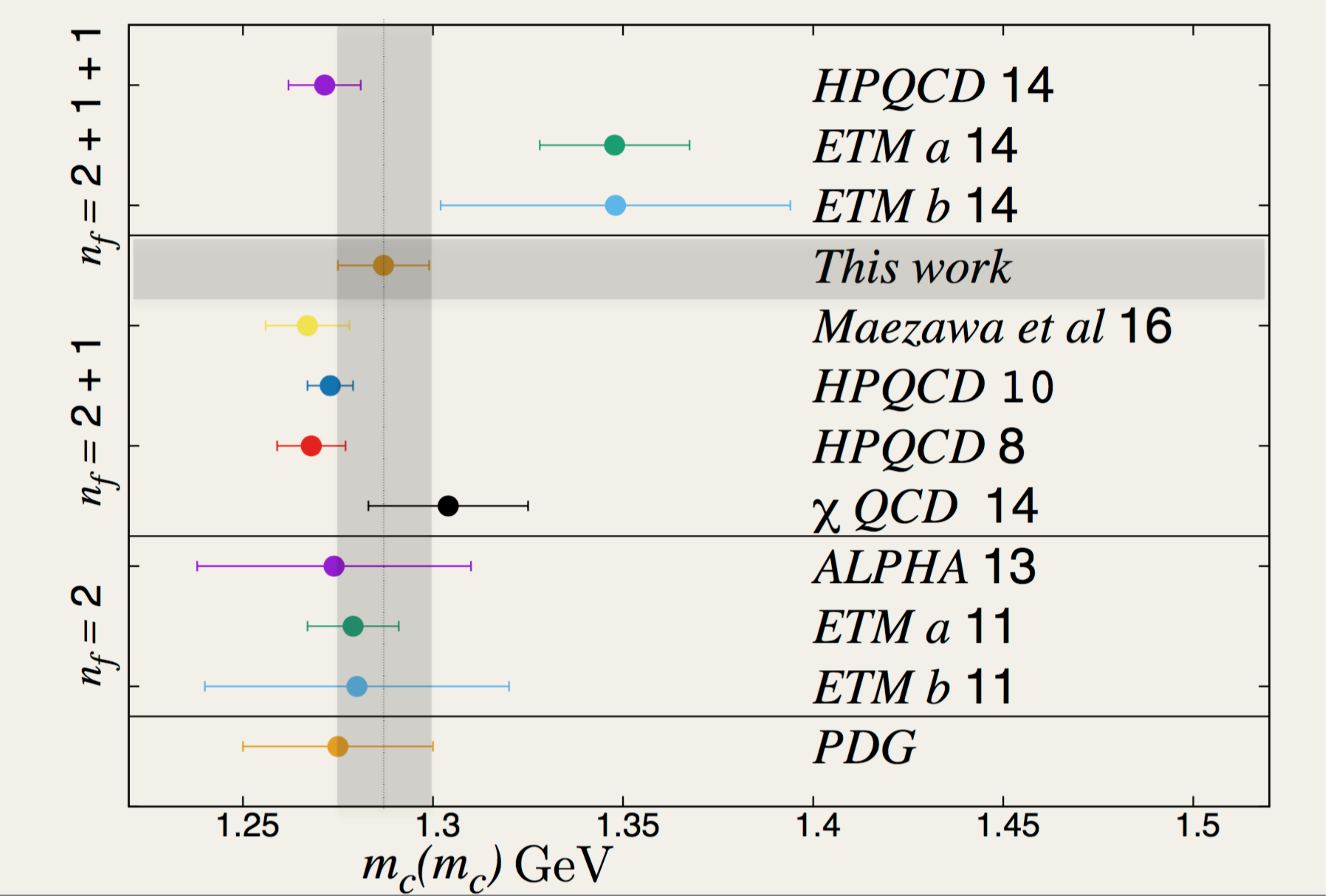}
    \\
  This topic was not discussed in Bologna.

\section{Search for new physics:}

    $D$-meson decays and $D$-mixing are places where new physics effects could lead to significant contributions. If the
    new degrees of freedom are heavy, i.e. of the order of the weak scale or larger, then we could use effective theories
    to integrate the new particles out and our theory tools would be in a similar good shape as in the $B$-system.
    The larger value of the strong coupling at the scale of charm quark mass will, however, still be a drawback, as well as the 
    uncertainty related to the detailed size of the standard model contribution. 
    On the other hand the new contributions could not be affected by the severe GIM cancellations present in the Standard Model
    part of $D$-mixing and some rare D-decays, leading thus to interesting bounds on the new couplings.
    New physics effects to $D$-meson decays were discussed in \cite{CHARM24,CHARM25}. 
    There is also some ongoing interest in finding bounds on the Yukawa-couplings  of the charm quark
    \cite{Perez:2015aoa,Bishara:2016jga,Botella:2016krk,Brivio:2015fxa,Koenig:2015pha}.
    The current bounds on the Higgs couplings were e.g. presented by Joachim Brod at BEAUTY 2016
    \\
    \includegraphics[width = 0.9 \textwidth]{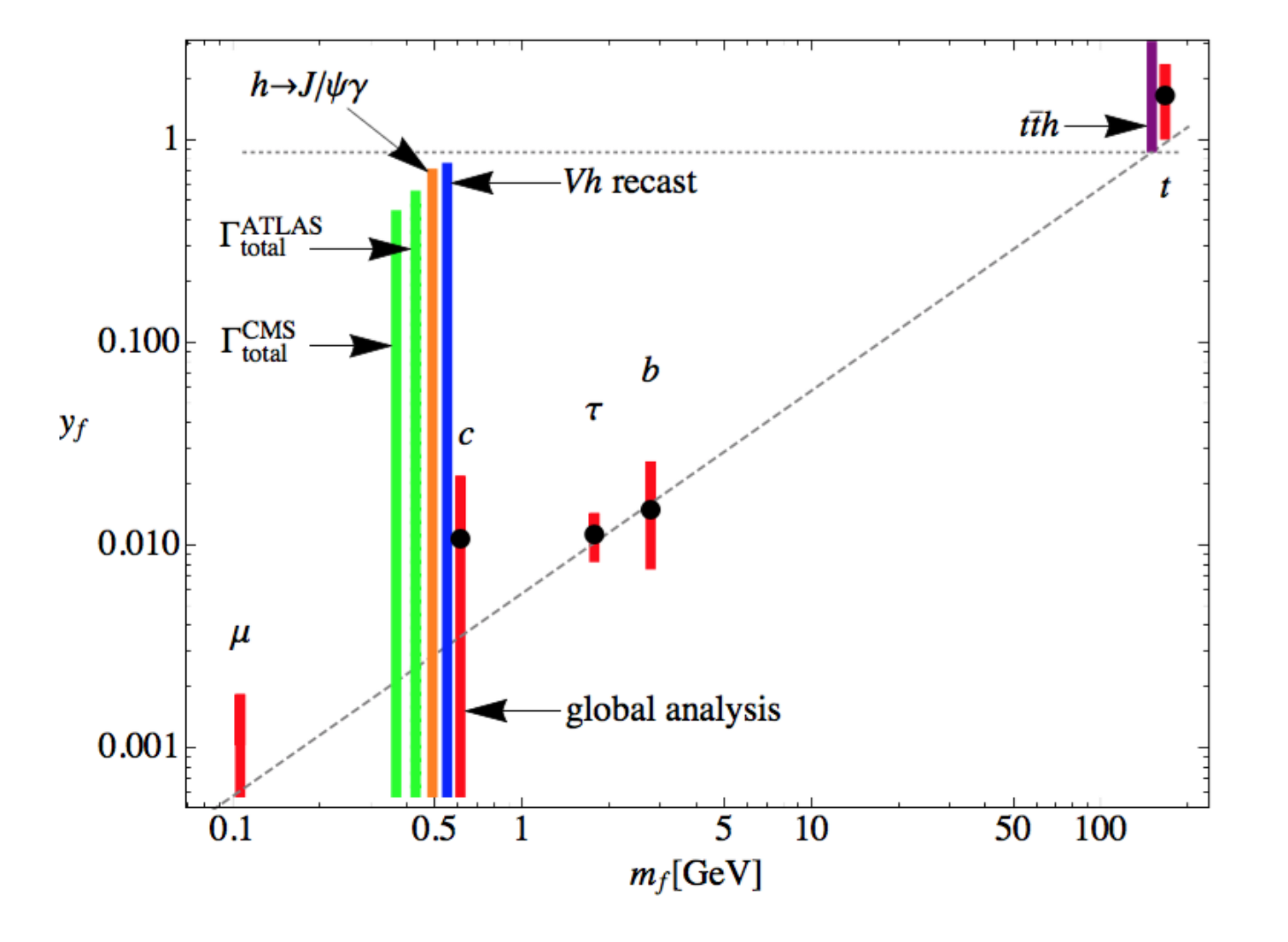}
    \\
    Here the charm coupling is severely unconstrained and leaves plenty of room for new effects.
    Indirect charm contributions are also crucial for precise Standard Model values of quantities
    that are very sensitive to new effects and there we currently find deviations of the order of 3 standard deviations, e.g. the
    anomalous magnetic moment of the muon $g-2$ and indirect CP-violation in Kaon-mixing, denoted by $\epsilon_K$, 
    see e.g. \cite{Sachrajda,Feng}.

\section{Conclusion}
    Further indications for the fact that charm phenomenology is a very rich field can be found in all the contributions
    to CHARM 2016, including the experimental summary \cite{CHARM40}. 
    \\
 We concentrated in these proceedings on the applicability of our theory tools to the charm sector.
    The seemingly obvious failure of the HQE for charm mixing, might also have different sources than a simple inapplicability:
    \begin{itemize}
    \item Quark hadron duality violating effects as low as 20 $\%$ could be the source of the discrepancy.
          Thus for other observables the HQE might still give decent estimates.
    \item Higher order terms in the HQE might be the dominant contribution due to a lifting of the GIM cancellations. Here
          again the HQE might still give good estimates for quantities that are not affected by severe GIM cancellations.
    \item Finally it is amusing to note, that it is still not completely excluded that we already observed new physics 
          effects in $D$ mixing.
    \end{itemize}
    The question whether the HQE gives reasonable estimates for charm observables that are not affected
    by strong GIM cancellations, can be well tested by theoretical studies of the $D$-meson lifetimes. The only missing 
    theoretical ingredients for this test are matrix elements of 4-quark operators, that can be determined with current 
    lattice technology.
    By the time of  CHARM 2018 in Novosibirsk \cite{CHARM41} 
    we will hopefully be closer to an answer of these questions.



\begin{thebibliography}{99}

\bibitem{CHARM01} Svjetlana Fajfer, these proceedings; talk at CHARM 2016.

\bibitem{CHARM02} Marco Gersabeck, these proceedings; talk at CHARM 2016.

\bibitem{HQE}
 V.~A.~Khoze, M.~A.~Shifman, N.~G.~Uraltsev and M.~B.~Voloshin,
  Sov.\ J.\ Nucl.\ Phys.\  {\bf 46} (1987) 112
   [Yad.\ Fiz.\  {\bf 46} (1987) 181].
 \\
  M.~A.~Shifman and M.~B.~Voloshin,
  Sov.\ Phys.\ JETP {\bf 64} (1986) 698
   [Zh.\ Eksp.\ Teor.\ Fiz.\  {\bf 91} (1986) 1180].
  \\
  I.~I.~Y.~Bigi and N.~G.~Uraltsev,
  Phys.\ Lett.\ B {\bf 280} (1992) 271.
  doi:10.1016/0370-2693(92)90066-D
  \\
  I.~I.~Y.~Bigi, N.~G.~Uraltsev and A.~I.~Vainshtein,
  Phys.\ Lett.\ B {\bf 293} (1992) 430
   Erratum: [Phys.\ Lett.\ B {\bf 297} (1992) 477]
  doi:10.1016/0370-2693(92)90908-M, 10.1016/0370-2693(92)91287-J
  [hep-ph/9207214].
  \\ 
  B.~Blok and M.~A.~Shifman,
  Nucl.\ Phys.\ B {\bf 399} (1993) 441
  doi:10.1016/0550-3213(93)90504-I
  [hep-ph/9207236].
  \\
  B.~Blok and M.~A.~Shifman,
  Nucl.\ Phys.\ B {\bf 399} (1993) 459
  doi:10.1016/0550-3213(93)90505-J
  [hep-ph/9209289].
  \\ 
  M.~Beneke, G.~Buchalla, C.~Greub, A.~Lenz and U.~Nierste,
  Phys.\ Lett.\ B {\bf 459} (1999) 631
  doi:10.1016/S0370-2693(99)00684-X
  [hep-ph/9808385].


\bibitem{HQEreview}
 A.~Lenz,
  Int.\ J.\ Mod.\ Phys.\ A {\bf 30} (2015) no.10,  1543005.
  doi:10.1142/S0217751X15430058
  \\
  A.~Lenz,
  arXiv:1405.3601 [hep-ph].

\bibitem{QCDfac}
  M.~Beneke, G.~Buchalla, M.~Neubert and C.~T.~Sachrajda,
  Phys.\ Rev.\ Lett.\  {\bf 83} (1999) 1914
  doi:10.1103/PhysRevLett.83.1914
  [hep-ph/9905312].
  \\
  M.~Beneke, G.~Buchalla, M.~Neubert and C.~T.~Sachrajda,
  Nucl.\ Phys.\ B {\bf 591} (2000) 313
  doi:10.1016/S0550-3213(00)00559-9
  [hep-ph/0006124].
  \\
  M.~Beneke, G.~Buchalla, M.~Neubert and C.~T.~Sachrajda,
  Nucl.\ Phys.\ B {\bf 606} (2001) 245
  doi:10.1016/S0550-3213(01)00251-6
  [hep-ph/0104110].
  \\
  M.~Beneke and M.~Neubert,
  Nucl.\ Phys.\ B {\bf 675} (2003) 333
  doi:10.1016/j.nuclphysb.2003.09.026
  [hep-ph/0308039].

\bibitem{GIM}
  S.~L.~Glashow, J.~Iliopoulos and L.~Maiani,
  Phys.\ Rev.\ D {\bf 2} (1970) 1285.
  doi:10.1103/PhysRevD.2.1285

\bibitem{Gersabeck:2015wna}
  M.~Gersabeck,
  PoS FWNP {\bf } (2015) 001
  [arXiv:1503.00032 [hep-ex]].
  \\
  M.~Gersabeck,
  Mod.\ Phys.\ Lett.\ A {\bf 27} (2012) 1230026
  doi:10.1142/S0217732312300261
  [arXiv:1207.2195 [hep-ex]].
  \\
   I.~Garzia [BESIII Collaboration],
  J.\ Phys.\ Conf.\ Ser.\  {\bf 631} (2015) no.1,  012043.
  doi:10.1088/1742-6596/631/1/012043





\bibitem{CHARM03} Alan Schwartz, these proceedings; talk at CHARM 2016.

\bibitem{CHARM04} Chris Parkes, these proceedings; talk at CHARM 2016.

\bibitem{CHARM05} Xiaoyan Shen, these proceedings; talk at CHARM 2016.

\bibitem{CHARM06} Gianluigi Boca, these proceedings; talk at CHARM 2016.


\bibitem{Lenz:2013pwa}
  A.~Lenz,
  arXiv:1311.6447 [hep-ph].

\bibitem{Aaij:2016phn}
  R.~Aaij {\it et al.} [LHCb Collaboration],
  Phys.\ Rev.\ Lett.\  {\bf 117} (2016) no.8,  082002
  doi:10.1103/PhysRevLett.117.082002
  [arXiv:1604.05708 [hep-ex]].


\bibitem{Aaij:2016ymb}
  R.~Aaij {\it et al.} [LHCb Collaboration],
  Phys.\ Rev.\ Lett.\  {\bf 117} (2016) no.8,  082003
   Addendum: [Phys.\ Rev.\ Lett.\  {\bf 117} (2016) no.10,  109902]
  doi:10.1103/PhysRevLett.117.082003, 10.1103/PhysRevLett.117.109902
  [arXiv:1606.06999 [hep-ex]].






\bibitem{CHARM07} Alfredo Valcarce, these proceedings; talk at CHARM 2016.

\bibitem{CHARM08} Pedro Gonzalez, these proceedings; talk at CHARM 2016.

\bibitem{CHARM09} Graham Moir, these proceedings; talk at CHARM 2016.

\bibitem{CHARM10} Timothy Burns, these proceedings; talk at CHARM 2016.

\bibitem{CHARM11} Francisco Fernandez, these proceedings; talk at CHARM 2016.

\bibitem{CHARM12} Gavin Cheung, these proceedings; talk at CHARM 2016.

\bibitem{CHARM13} Alessandro Pilloni, these proceedings; talk at CHARM 2016.

\bibitem{CHARM14} Sinead Ryan, these proceedings; talk at CHARM 2016.

\bibitem{CHARM15} Nora Brambillak, these proceedings; talk at CHARM 2016.



\bibitem{CHARM16} Frank Geurts, these proceedings; talk at CHARM 2016.

\bibitem{CHARM17} Francois Arleo, these proceedings; talk at CHARM 2016.

\bibitem{CHARM18} Andrea Beraudo, these proceedings; talk at CHARM 2016.

\bibitem{CHARM19} Antonio Vairo, these proceedings; talk at CHARM 2016.

\bibitem{CHARM20} Martin Cleven, these proceedings; talk at CHARM 2016.

       
\bibitem{CHARM21} Johann Haidenbauer, these proceedings; talk at CHARM 2016.

\bibitem{CHARM22} Wei Wang, these proceedings; talk at CHARM 2016.

\bibitem{CHARM23} Aida El-Khadra, these proceedings; talk at CHARM 2016.

\bibitem{CHARM24} Nejc Kosnik, these proceedings; talk at CHARM 2016.

\bibitem{CHARM25} Ayan Paul, these proceedings; talk at CHARM 2016.

\bibitem{Perez:2015aoa}
  G.~Perez, Y.~Soreq, E.~Stamou and K.~Tobioka,
  Phys.\ Rev.\ D {\bf 92} (2015) no.3,  033016
  doi:10.1103/PhysRevD.92.033016
  [arXiv:1503.00290 [hep-ph]].

\bibitem{Bishara:2016jga}
  F.~Bishara, U.~Haisch, P.~F.~Monni and E.~Re,
  arXiv:1606.09253 [hep-ph].

\bibitem{Botella:2016krk}
  F.~J.~Botella, G.~C.~Branco, M.~N.~Rebelo and J.~I.~Silva-Marcos,
  arXiv:1602.08011 [hep-ph].

\bibitem{Brivio:2015fxa}
  I.~Brivio, F.~Goertz and G.~Isidori,
  Phys.\ Rev.\ Lett.\  {\bf 115} (2015) no.21,  211801
  doi:10.1103/PhysRevLett.115.211801
  [arXiv:1507.02916 [hep-ph]].

\bibitem{Koenig:2015pha}
  M.~König and M.~Neubert,
  JHEP {\bf 1508} (2015) 012
  doi:10.1007/JHEP08(2015)012
  [arXiv:1505.03870 [hep-ph]].


\bibitem{CHARM26} Jonas Rademacker, these proceedings; talk at CHARM 2016.
          
\bibitem{CHARM27} Benoit Loiseau, these proceedings; talk at CHARM 2016.
         
\bibitem{CHARM28} Satoshi Nakamura, these proceedings; talk at CHARM 2016.
          
\bibitem{CHARM29} Patricia Magalhaes, these proceedings; talk at CHARM 2016.
          
\bibitem{CHARM30} Pietro Santorelli, these proceedings; talk at CHARM 2016.

          
\bibitem{CHARM31} Guido Martinelli, these proceedings; talk at CHARM 2016.
         
\bibitem{CHARM32} Marco Ciuchinik, these proceedings; talk at CHARM 2016.
         
\bibitem{CHARM33} Lorenza Riggio, these proceedings; talk at CHARM 2016.


\bibitem{Lenz:2013aua}
  A.~Lenz and T.~Rauh,
  Phys.\ Rev.\ D {\bf 88} (2013) 034004
  doi:10.1103/PhysRevD.88.034004
  [arXiv:1305.3588 [hep-ph]].


           
\bibitem{CHARM34} Denis Derkach, these proceedings; talk at CHARM 2016.


\bibitem{charmDM} T.Jubb, M. Kirk and A. Lenz in preparation.

          
\bibitem{Wittig}
Hartmut Wittig, talk at LATTICE 2016.


\bibitem{Sachrajda}
Chris Sachrajda, talk at KAON 2016.

\bibitem{Feng}
Xu Feng, talk at KAON 2016.


 \bibitem{CHARM35} Roy Briere, these proceedings; talk at CHARM 2016.

\bibitem{Maezawa:2016vgv}
  Y.~Maezawa and P.~Petreczky,
  Phys.\ Rev.\ D {\bf 94} (2016) no.3,  034507
  doi:10.1103/PhysRevD.94.034507
  [arXiv:1606.08798 [hep-lat]].

\bibitem{Lenz:2012mb}
  A.~Lenz,
  arXiv:1205.1444 [hep-ph].

\bibitem{Amhis:2014hma}
  web-update of 
  Y.~Amhis {\it et al.} [Heavy Flavor Averaging Group (HFAG) Collaboration],
  arXiv:1412.7515 [hep-ex].

\bibitem{Artuso:2015swg}
  M.~Artuso, G.~Borissov and A.~Lenz,
  arXiv:1511.09466 [hep-ph].

\bibitem{Lenz:2006hd}
  A.~Lenz and U.~Nierste,
  JHEP {\bf 0706} (2007) 072
  doi:10.1088/1126-6708/2007/06/072
  [hep-ph/0612167].

\bibitem{Beneke:2003az}
  M.~Beneke, G.~Buchalla, A.~Lenz and U.~Nierste,
  Phys.\ Lett.\ B {\bf 576} (2003) 173
  doi:10.1016/j.physletb.2003.09.089
  [hep-ph/0307344].

\bibitem{Ciuchini:2003ww}
  M.~Ciuchini, E.~Franco, V.~Lubicz, F.~Mescia and C.~Tarantino,
  JHEP {\bf 0308} (2003) 031
  doi:10.1088/1126-6708/2003/08/031
  [hep-ph/0308029].

\bibitem{Beneke:2002rj}
  M.~Beneke, G.~Buchalla, C.~Greub, A.~Lenz and U.~Nierste,
  Nucl.\ Phys.\ B {\bf 639} (2002) 389
  doi:10.1016/S0550-3213(02)00561-8
  [hep-ph/0202106].

\bibitem{Dighe:2001gc}
  A.~S.~Dighe, T.~Hurth, C.~S.~Kim and T.~Yoshikawa,
  Nucl.\ Phys.\ B {\bf 624} (2002) 377
  doi:10.1016/S0550-3213(01)00655-1
  [hep-ph/0109088].


\bibitem{Beneke:1998sy}
  M.~Beneke, G.~Buchalla, C.~Greub, A.~Lenz and U.~Nierste,
  Phys.\ Lett.\ B {\bf 459} (1999) 631
  doi:10.1016/S0370-2693(99)00684-X
  [hep-ph/9808385].

\bibitem{Beneke:1996gn}
  M.~Beneke, G.~Buchalla and I.~Dunietz,
  Phys.\ Rev.\ D {\bf 54} (1996) 4419
   Erratum: [Phys.\ Rev.\ D {\bf 83} (2011) 119902]
  doi:10.1103/PhysRevD.54.4419, 10.1103/PhysRevD.83.119902
  [hep-ph/9605259].

\bibitem{Bobrowski:2010xg}
  M.~Bobrowski, A.~Lenz, J.~Riedl and J.~Rohrwild,
  JHEP {\bf 1003} (2010) 009
  doi:10.1007/JHEP03(2010)009
  [arXiv:1002.4794 [hep-ph]].


\bibitem{Rosner:2015wva}
  J.~L.~Rosner, S.~Stone and R.~S.~Van de Water,
  [arXiv:1509.02220 [hep-ph]].

\bibitem{Gottlieb} Steven Gottlieb, talk at LATTICE 2016.




\bibitem{Muller:2015lua}
  S.~Müller, U.~Nierste and S.~Schacht,
  Phys.\ Rev.\ D {\bf 92} (2015) no.1,  014004
  doi:10.1103/PhysRevD.92.014004
  [arXiv:1503.06759 [hep-ph]].

\bibitem{Gronau:2015rda}
  M.~Gronau,
  Phys.\ Rev.\ D {\bf 91} (2015) no.7,  076007
  doi:10.1103/PhysRevD.91.076007
  [arXiv:1501.03272 [hep-ph]].

\bibitem{Gao:2014ena}
  D.~N.~Gao,
  Phys.\ Rev.\ D {\bf 91} (2015) no.1,  014019
  doi:10.1103/PhysRevD.91.014019
  [arXiv:1411.0768 [hep-ph]].

\bibitem{Jung:2014jfa}
  M.~Jung and S.~Schacht,
  Phys.\ Rev.\ D {\bf 91} (2015) no.3,  034027
  doi:10.1103/PhysRevD.91.034027
  [arXiv:1410.8396 [hep-ph]].

\bibitem{Santorelli:2014kva}
  P.~Santorelli,
  EPJ Web Conf.\  {\bf 80} (2014) 00051
  doi:10.1051/epjconf/20148000051
  [arXiv:1410.1719 [hep-ph]].

\bibitem{Hiller:2013awa}
  G.~Hiller, M.~Jung and S.~Schacht,
  PoS EPS {\bf -HEP2013} (2013) 371
  [arXiv:1311.3883 [hep-ph]].


\bibitem{Buccella:2013tya}
  F.~Buccella, M.~Lusignoli, A.~Pugliese and P.~Santorelli,
  Phys.\ Rev.\ D {\bf 88} (2013) no.7,  074011
  doi:10.1103/PhysRevD.88.074011
  [arXiv:1305.7343 [hep-ph]].


\bibitem{Hansen:2012tf}
  M.~T.~Hansen and S.~R.~Sharpe,
  Phys.\ Rev.\ D {\bf 86} (2012) 016007
  doi:10.1103/PhysRevD.86.016007
  [arXiv:1204.0826 [hep-lat]].

\bibitem{Carrasco:2015pra}
  N.~Carrasco {\it et al.} [ETM Collaboration],
  Phys.\ Rev.\ D {\bf 92} (2015) no.3,  034516
  doi:10.1103/PhysRevD.92.034516
  [arXiv:1505.06639 [hep-lat]].

 \bibitem{Kronfeld} Andreas Kronfeld, talk at LATTICE 2016.


\bibitem{Falk:2001hx}
  A.~F.~Falk, Y.~Grossman, Z.~Ligeti and A.~A.~Petrov,
  Phys.\ Rev.\ D {\bf 65} (2002) 054034
  doi:10.1103/PhysRevD.65.054034
  [hep-ph/0110317].

\bibitem{Falk:2004wg}
  A.~F.~Falk, Y.~Grossman, Z.~Ligeti, Y.~Nir and A.~A.~Petrov,
  Phys.\ Rev.\ D {\bf 69} (2004) 114021
  doi:10.1103/PhysRevD.69.114021
  [hep-ph/0402204].

\bibitem{Jubb:2016mvq}
  T.~Jubb, M.~Kirk, A.~Lenz and G.~Tetlalmatzi-Xolocotzi,
  arXiv:1603.07770 [hep-ph].




\bibitem{Georgi:1992as}
  H.~Georgi,
  Phys.\ Lett.\ B {\bf 297} (1992) 353
  doi:10.1016/0370-2693(92)91274-D
  [hep-ph/9209291].


\bibitem{Ohl:1992sr}
  T.~Ohl, G.~Ricciardi and E.~H.~Simmons,
  Nucl.\ Phys.\ B {\bf 403} (1993) 605
  doi:10.1016/0550-3213(93)90364-U
  [hep-ph/9301212].

\bibitem{Bigi:2000wn}
  I.~I.~Y.~Bigi and N.~G.~Uraltsev,
  Nucl.\ Phys.\ B {\bf 592} (2001) 92
  doi:10.1016/S0550-3213(00)00604-0
  [hep-ph/0005089].


\bibitem{Bobrowski:2012jf}
  M.~Bobrowski, A.~Lenz and T.~Rauh,
  arXiv:1208.6438 [hep-ph].

\bibitem{PDG}
  C.~Patrignani,
  Chin.\ Phys.\ C {\bf 40} (2016) no.10,  100001.
  doi:10.1088/1674-1137/40/10/100001


\bibitem{mc}
Katsumasa Nakayama, talk at LATTICE 2016. 

\bibitem{CHARM40} Angelo di Canto, these proceedings; talk at CHARM 2016.

\bibitem{CHARM41} Alexander Bondar, these proceedings; talk at CHARM 2016.


\end{thebibliography}
\end{document}